\documentclass[twocolumn,twoside,slac_two]{revtex4}

\usepackage[]{graphicx}
\usepackage{amsmath,amssymb}

\usepackage{fancyhdr}
\newcommand{\E}{E_{0}}
\newcommand{\Ethr}{\varepsilon_{\text{thr}}}
\newcommand{\rhom}{\rho_{\mu}}
\newcommand{\rhoc}{\rho_{ch}}
\newcommand{\xmax}{x_{\text{max}}}
\newcommand{\xgrnd}{x_{0}}
\newcommand{\depth}{g/cm$^{2}$}

\begin{document}

\title{The Relation Between Charged Particles and Muons With Threshold Energy 1 GeV in Extensive Air Showers Registered at the Yakutsk EAS Array}

\author{S.~P.~Knurenko}
\author{A.~K.~Makarov}
\author{M.~I.~Pravdin}
\author{A.~Sabourov}

\affiliation{Yu. G. Shafer Institute of cosmophysical research and aeronomy SB RAS}

\begin{abstract}
  Characteristics of the  muon component in EAS are analyzed together with their fluctuations. The aim of this analysis~--- a comparison of experimental data with computational results obtained within frameworks of various hadron interaction models for protons and iron nuclei and an estimation of cosmic ray mass composition in the ultra-high energy region.
\end{abstract}

\maketitle

\thispagestyle{fancy}

\section{Introduction}

The Yakutsk complex array for many years measures three main observables of extensive air showers (EAS): total charged component, muons with $\Ethr \ge 1$~GeV and Cherenkov light \citep{bib1}. Using these data we estimated the EAS energy with model independent quasi-calorimetric method \citep{bib2} and determined the depth of maximum shower development (by the measured Cherenkov light lateral distribution, using the parameter $p = \lg{Q(200) / Q(550)}$ and by the shape of the Cherenkov light pulse, $\tau_{1/2}$) \citep{bib3, bib4}. The relative muon content at different core distances was measured \citep{bib5, bib6} and the cosmic ray (CR) mass composition was estimated by various EAS characteristics~\citep{bib7, bib8, bib9}.

In this paper we analyze the muon component of EAS: mean characteristics, muon content and its fluctuations at fixed energy. The analysis is conducted within the framework of QGSJet~II \citep{bib10} and EPOS \citep{bib11} hadron interaction models involving computations for primary particles of different masses using CORSIKA-6.900 code \citep{bib12}.

\section{Muon lateral distribution function}

\begin{figure}
  \centering
  \includegraphics[width=0.49\textwidth, clip]{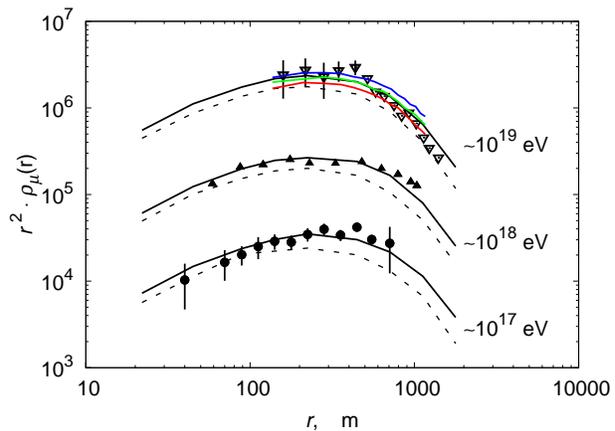}
  \caption{Lateral distribution of charged particles for fixed energy values $10^{17}, 10^{28}, 10^{19}$~eV and zenith angle $\theta = 15^{\circ}$. Symbols refer to the Yakutsk experiment. Solid lines denote computational results with QGSJet+FLUKA \citep{bib13} for protons, dotted line~--- for iron, red line~--- EPOS+UrQMD \citep{bib14} for proton, green~--- carbon, blue~--- iron.}
  \label{fig1}
\end{figure}

In Figure~\ref{fig1}  examples of mean lateral distributions for muons at different energies are displayed. The muon lateral distribution function (LDF) is significantly lower then that of the charged component and can be effectively measured in individual events at $\E \ge 10^{17}$~eV within the core distance range $100-800$~m. Thus, as a classification parameter in this energy region, a parameter $\rhom(600)$ could be used~--- the density of muon flux at $600$~m from shower core.

Solid and dotted lines on the figure denote computational results obtained with QGSJet(UrQMD) models for proton (solid) and iron (dotted). It is seen from  Figure~\ref{fig1} that the muon LDF from protons is  steeper than that from iron nuclei and this difference is especially pronounced at large core distances. Qualitative comparison of computational results with the experiment reveals a better agreement with a heavier component of primary CR at $\E \le 10^{18}$~eV and with lighter at $\E \sim 10^{19}$~eV. This feature could be stressed out if one puts parameter $r^{2} \cdot \rho(r)$ on the $y$-axis of a plot instead of simple $\rho(r)$.

\section{Muon portion and its dependence on angle, energy and the depth of maximum EAS development}

We considered the dependence of $\rhom/\rhoc$ on the length of shower development after the maximum~--- $\Delta\lambda = \xgrnd / \cos{\theta} - \xmax$. In highly inclined showers the muon content increases proportionally to $\xgrnd/\cos{\theta}$ value, where $\xgrnd = 1020$~\depth  for Yakutsk.

It is a known fact that the depth of maximum EAS development differs significantly, depending on the kind of primary particle and, hence, this feature could be used in the analysis of the CR mass composition: for instance, by fixing the $\Delta\lambda$ parameter and studying the fluctuations of $\rhom/\rhoc$ value. This technique is rather similar to one proposed by \citet{bib15}.

Shower parameters calculated with CORSIKA code were modified by applying distortions according to experimental errors. Parameters measured in experiment (e.g. $\cos{\theta}, \xmax, \rhoc(r), \rhom(r)$) for every shower were rolled with the normal distribution with $\sigma$ parameter according to the experiment:

\begin{displaymath}
  \begin{aligned}
    \sigma(\theta) &= 3 \cdot \sec{\theta};\\
    \sigma(\xmax) &= 40\text{g/cm}^{2};\\
    \sigma(\rho_{r}) &= \sqrt{\rho_{r}^{2} \cdot \left(0.025 + \frac{1.2}{s_{\text{det}} \cdot \rho_{r} \cdot \cos{\theta}}\right)}
    \end{aligned}
\end{displaymath}
where $s_{\text{det}}$ is the area of the detector.

\begin{figure*}
  \centering
  \includegraphics[width=0.49\textwidth, clip]{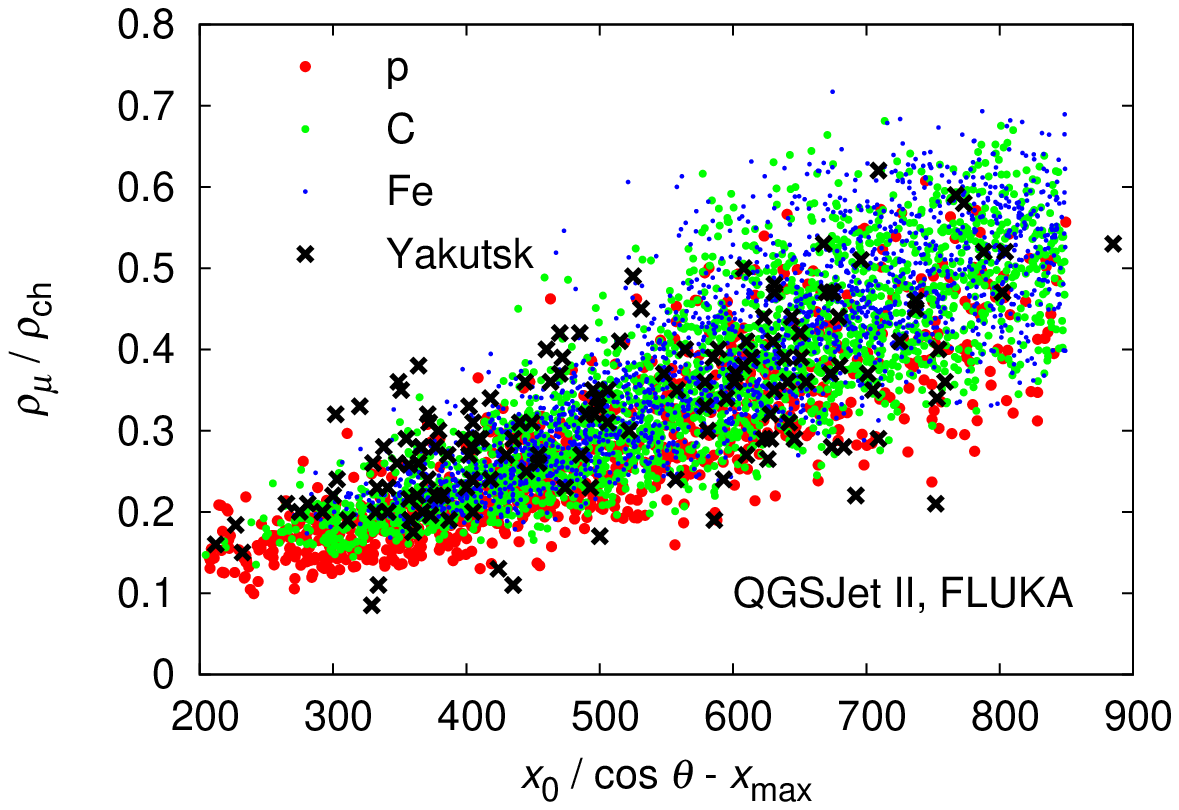}
  \includegraphics[width=0.49\textwidth, clip]{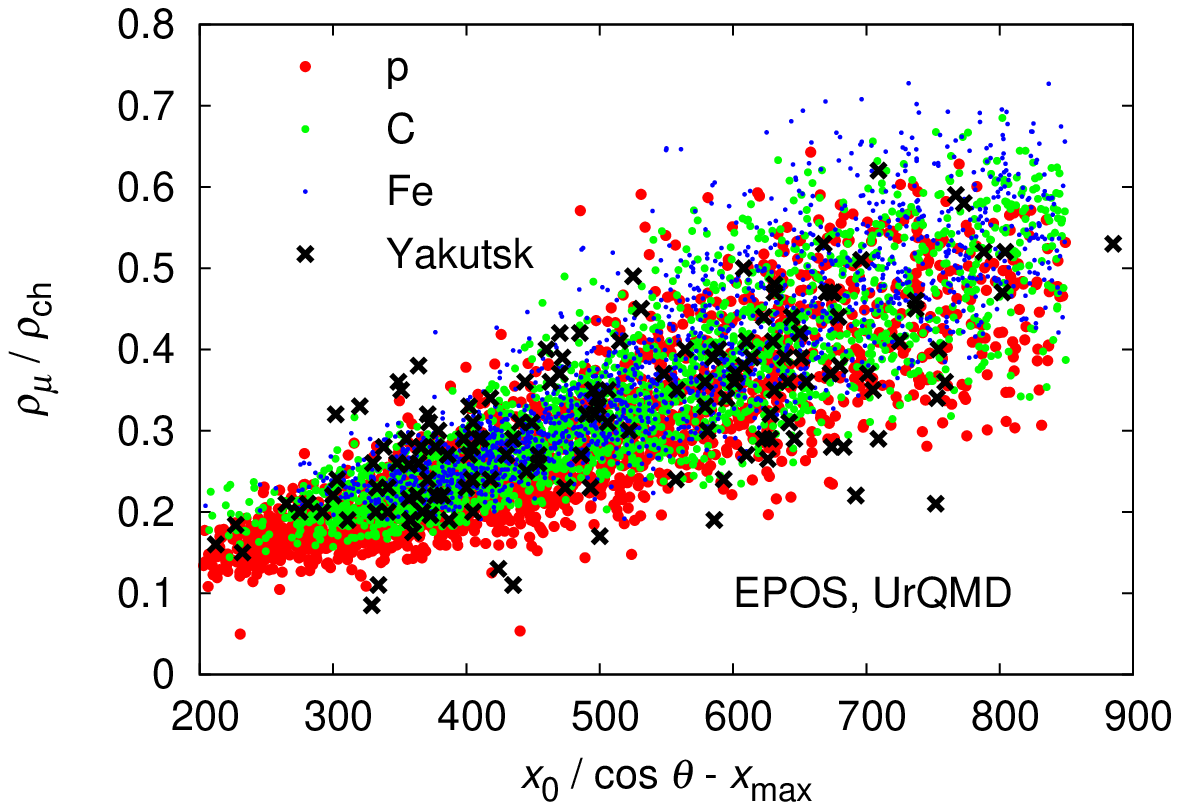}
  \caption{A dependence of muon portion with $\Ethr \ge 1$~GeV on the length of track in the atmosphere for individual showers with $\theta = 0 - 50^{\circ}$ and energy $10^{18}$~eV. On the left~--- results obtained with QGSJet~II model, on the right~--- with EPOS model.}
  \label{fig2}
\end{figure*}

Figure~\ref{fig2} shows the dependence of $\rhom/\rhoc$ on the length of cascade development after the shower maximum compared with computational results. A strong correlation is observed between the muon content and the length of track in the atmosphere after the shower maximum. It is also seen that experimental data are in good agreement with simulation results.

\section{Mean characteristics}

\begin{figure}
  \centering
  \includegraphics[width=0.45\textwidth]{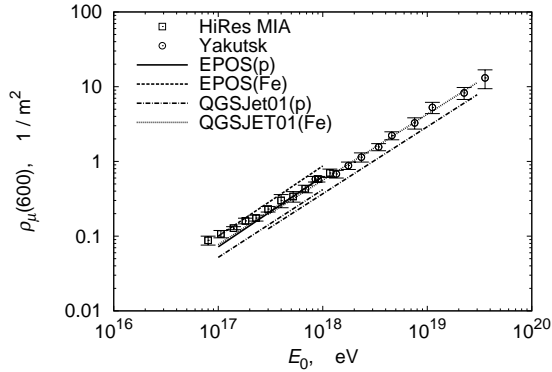}
  \caption{Muon density $\rhom(600)$ measured in Yakutsk experiment as a function of primary energy compared to model calculations.}
  \label{fig3}
\end{figure}

Figure~\ref{fig3} shows the energy dependence of $\rhom(600)$ obtained in the Yakutsk and MIA EAS experiments \citep{bib16}. A good agreement is observed between the two.

Computational results obtained with EPOS and QGSJet01 from the work by \citet{bib16} are denoted with lines, dotted line represent our simulations with QGSJet~II for protons and iron nuclei. A comparison of our computations with the results obtained by \citet{bib16} reveals that virtually there is no  difference between QGSJet01 and QGSJet~II. A significant discrepancy is observed between EPOS and QGSJet~II and it, as we believe, is associated with different amounts of muons generated by models. For example, the $\rhom(600)$ value calculated with EPOS for proton coincides with $\rhom(600)$ obtained with QGSJet~II for iron. Thus, a comparison of experimental data with model calculations result in controversial conclusions on CR mass composition. According to EPOS, at energies up to $2 \times 10^{17}$~eV CRs consist of iron nuclei and above that energy, up to $10^{19}$~eV~--- of protons. In the energy interval $10^{17} - 3 \times 10^{18}$~eV QGSJet~II computations agree with the experiment quite well if primary particles are iron nuclei and above $3 \times 10^{18}$~eV the mass composition might be mixed with portion of protons and helium nuclei not less than $50-60$\,\%. More precise estimation of CR mass composition could be derived after improvement of theoretical models and selecting one, that describes experimental EAS data better then others.

\begin{figure*}
  \centering
  \includegraphics[width=0.49\textwidth, clip]{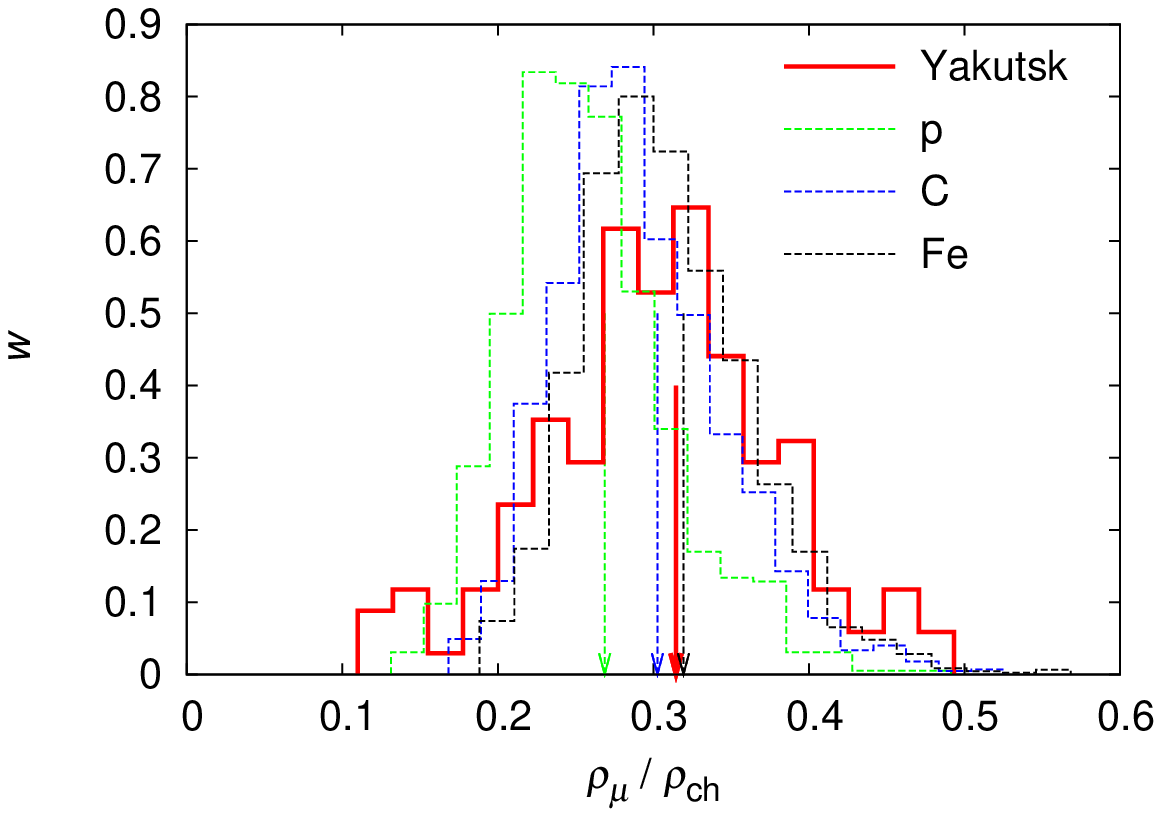}
  \includegraphics[width=0.49\textwidth, clip]{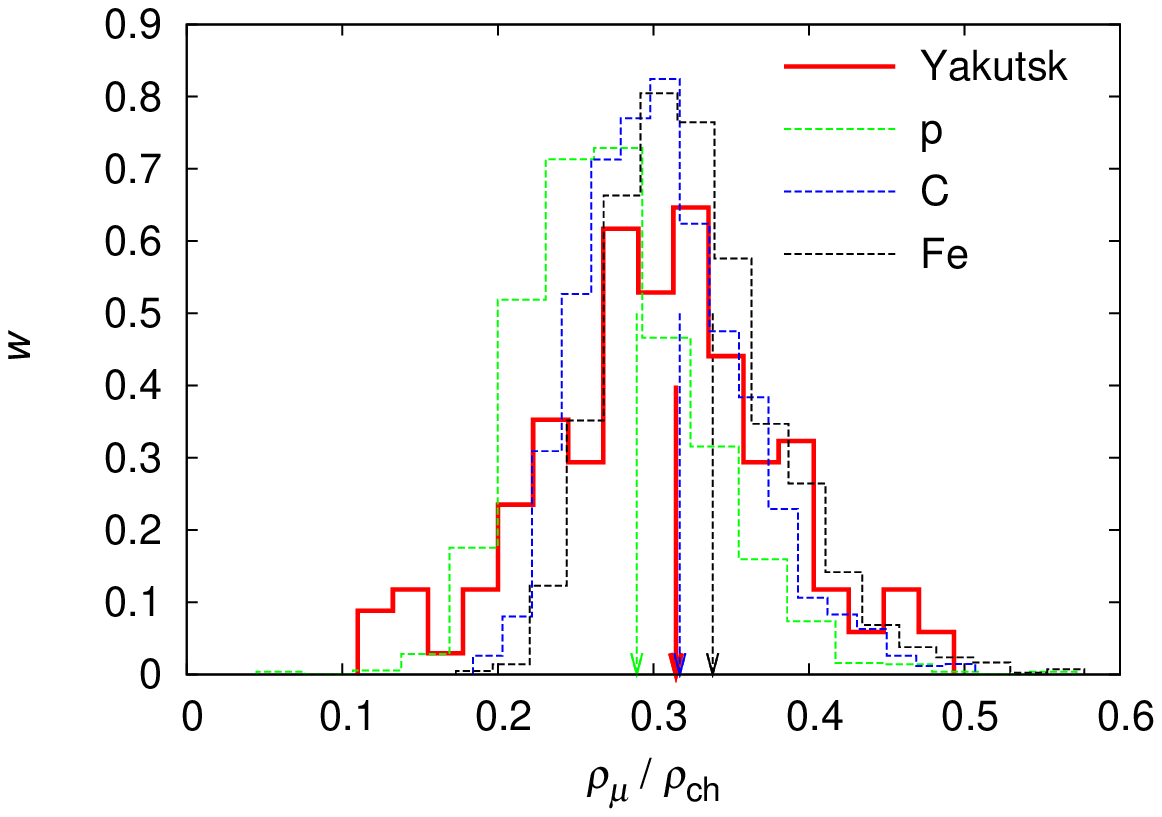}
  \caption{A distribution of the $\rhom/\rhoc$ relation, normalized to the track length $500$~\depth. On the left~--- according to models QGSJet~II(FLUKA), on the right~--- according to EPOS(UrQMD).}
  \label{fig4}
\end{figure*}

\section{Fluctuations of $\rhom/\rhoc$ relation on the ground level at the energy $\sim 10^{18}$~eV}

Showers initiated by different nuclei have differing altitudes of the maximum which in turn means that different numbers of muons are generated in these showers. It also means that they cover different paths in the atmosphere. By analyzing the tracks of muons that they pass in the atmosphere after the maximum of shower development we can try to estimate the composition of cosmic rays. With this aim in view, by choosing the mean zenith angle $36^{\circ}$ (which corresponds to the track length after the maximum $\Delta\lambda = 500$~\depth), let us normalize the values of muon content to this level and consider their fluctuations.

Results are presented in Figure~\ref{fig4}. Also shown are the computational results obtained with QGSJet~II and EPOS models for various nuclei. Measured values of fluctuations are presented in Table~\ref{tab1}. Obtained results have shown that within this method fluctuations of $\rhom(600)/\rhoc(600)$ parameters do not allow to estimate the CR mass composition. However, mean values from different nuclei differ. Besides, QGSJet~II hints at a heavier composition than that of EPOS: according to first one, the composition of selected showers shifts towards heavier nuclei; according to second one, showers correspond to nuclei of intermediate group. On the whole, both models argue for a mixed composition.

However, if one takes into account gamma-photons generated in ground covering muon detectors, the mean value of $\rhom(600)/\rhoc(600)$ relation decreases and the composition shifts towards lighter nuclei (protons-helium-carbon)~\cite{bib18}.
\begin{table*}
  \caption{Fluctuations of $\rhom/\rhoc$ relation, normalized to the length of track $500$~\depth}
  \label{tab1}
  \centering
  \begin{tabular}{ccccccccc}
    \hline
    && \multicolumn{3}{c}{QGSJet~II, FLUKA} & \multicolumn{3}{c}{EPOS, UrQMD} \\
    &Yakutsk & Yakutsk\footnote{With respect to contribution from gammas generated in the shielding of detector (Dedenko, 2010)}&p & C & Fe & p & C & Fe \\
    \hline
    $\left< \rho_{\mu} / \rho_{\text{ch}} \right>$ &
    $0.3145$ & $0.2768$ & $0.2687$ & $0.3025$ & $0.3193$ & $0.2893$ & $0.3170$ & $0.3381$ \\
    \hline
    $\sigma$ &
    $0.0747$ & $0.0657$ & $0.0517$ & $0.0541$ & $0.0536$ & $0.0563$ & $0.0511$ & $0.0539$ \\
    \hline
  \end{tabular}
\end{table*}

\section{Conclusions}

Within the framework of QGSJet~II and EPOS hadron interaction models using the CORSIKA code the values of muon portion $\rhom/\rhoc$ at core distance $r = 600$~m were obtained. A relation between the muon portion and a distance to the depth of shower maximum $\Delta\lambda$ was also obtained. A comparison of the dependency with  experiment has shown that taking account of the experimental errors in the simulation data, a good agreement is observed between simulation and experiment.

A comparison of the muon portion distribution with computational results points towards a mixed cosmic ray composition near $\E \ge 10^{18}$~eV. Large fluctuations of the muon portion prevent revealing of a single determined group of nuclei. A more detailed analysis is required, involving possible systematics of muon density measurement in the Yakutsk experiment.

\begin{acknowledgements}
  The work is supported by RFBR grants 08-02-00348-a, 09-02-12028 ofi-m and FANI g.k. 02.740.11.0248, 02.518.11.7173.
\end{acknowledgements}

\end{document}